\documentclass[letterpaper]{article} 
\usepackage{aaai24}  
\usepackage{times}  
\usepackage{helvet}  
\usepackage{courier}  
\usepackage[hyphens]{url}  
\usepackage{graphicx} 
\usepackage{natbib}  
\usepackage{caption} 
\usepackage{algorithm}
\usepackage{algorithmic}

\usepackage{newfloat}
\usepackage{listings}
\DeclareCaptionStyle{ruled}{labelfont=normalfont,labelsep=colon,strut=off} 
\floatstyle{ruled}
\newfloat{listing}{tb}{lst}{}
\floatname{listing}{Listing}
\usepackage{array, amsmath}
\usepackage{diagbox, multirow}
\usepackage[caption=false]{subfig}
\DeclareMathSizes{9.5}{8.5}{7.5}{6.5}
\usepackage{xcolor} 

\title{From GARCH to Neural Network for Volatility Forecast}
\author{
    Pengfei Zhao\equalcontrib\textsuperscript{\rm 1},
    Haoren Zhu\equalcontrib\textsuperscript{\rm 2},
    Wilfred Siu Hung NG\textsuperscript{\rm 2},
    Dik Lun Lee \textsuperscript{\rm 2}}
\affiliations {
    \textsuperscript{\rm 1} Guangdong Provincial Key Laboratory of Interdisciplinary Research and Application for Data Science, BNU-HKBU United International College,\\
    \textsuperscript{\rm 2}Hong Kong University of Science and Technology,\\
    ericpfzhao@uic.edu.cn,
    \{hzhual, wilfred, dlee\}@cse.ust.hk,    
}

\begin{document}

\maketitle

\begin{abstract}
Volatility, as a measure of uncertainty, plays a crucial role in numerous financial activities such as risk management. The Econometrics and Machine Learning communities have developed two distinct approaches for financial volatility forecasting: the stochastic approach and the neural network (NN) approach. Despite their individual strengths, these methodologies have conventionally evolved in separate research trajectories with little interaction between them. This study endeavors to bridge this gap by establishing an equivalence relationship between models of the GARCH family and their corresponding NN counterparts. With the equivalence relationship established, we introduce an innovative approach, named GARCH-NN, for constructing NN-based volatility models. It obtains the NN counterparts of GARCH models and integrates them as components into an established NN architecture, thereby seamlessly infusing volatility stylized facts (SFs) inherent in the GARCH models into the neural network. We develop the GARCH-LSTM model to showcase the power of the GARCH-NN approach. Experiment results validate that amalgamating the NN counterparts of the GARCH family models into established NN models leads to enhanced outcomes compared to employing the stochastic and NN models in isolation.
\end{abstract}

\section{Introduction}

In empirical finance, the volatility of asset returns is a measure of risk, which helps investors decide if the returns of assets justify the risks. Further, volatility is used by financial institutions to assess their risks using value-at-risk (VaR) models. The ability to forecast the volatility or variance of asset returns can help investors to anticipate risks and to reduce loss, which is important for risk management. Financial forecasting can be categorized into stochastic and machine learning (ML) approaches. Existing econometrics solutions fall into the stochastic category. For example, the famous \textit{generalized autoregressive conditional heteroscedasticity} (GARCH) family models for volatility modeling have been widely used in risk modeling for decades. Stochastic econometric time series models have the advantage that they can be theoretically described based on statistical logic. However, they rely on many assumptions (e.g., explanatory variables must be stationary), which may not align with the highly dynamic, nonlinear, and complex market reality. 

Recently, ML approaches have been increasingly used in volatility forecast \cite{DL-volatility-forecast-survey}. However, ML approaches are mostly uninterpretable and lack support from financial domain knowledge. While ML models have the power to approximate nonlinear functions and fewer restrictions, they have several limitations: (1) The complexity of NN models poses challenges in explaining the model's outputs, thus reducing confidence in the model's reliability. The issue is magnified in the financial sector where risks and responsibilities are at stake, compelling practitioners to understand how forecasts and decisions
are made to build up trust in the model. (2) There are a large number of models that fit the training data well, but few generalize well \cite{noise2001}. The challenge mainly comes from the non-stationarity and noisiness of real-world financial time series data. 
A small training timeframe may induce overfitting in NN models due to the presence of noise. Conversely, an excessively large timeframe can introduce inconsistency between the temporal relationship of inputs, confusing NN models during the learning process due to data non-stationarity.
Recent work \cite{zeng2022transformers} shows that simple one-layer linear models outperform sophisticated Transformer-based models on long-term time series forecasting. (3) Well-established NN models are often general ML solutions and the network structure is not designed to capture the characteristics of the volatility time series. Thus, statistical models like GARCH are still the dominant volatility forecast models until now due to their simplicity, high interpretability, and satisfactory performance.

Stochastic and NN solutions have complementary strengths, but existing studies treat them as separate research directions without much interaction in between. This paper aims to develop an approach that can exploit the well-established foundation of the GARCH family of volatility models in NN-based volatility models to improve prediction accuracy as well as enhance the interpretability of the NN models. In particular, econometricians have discovered the fact that volatility, unlike most market variables that remain largely unpredictable, has specific characteristics named ``stylized facts'' (SFs) \cite{Masset_StylizedFacts2011} (also referred to as ``volatility characteristics'' in the literature) that can increase the accuracy of volatility forecast. Examples of stylized facts are: the change of volatility tends to persist (\textit{volatility clustering}), to be higher in declining markets than in rising markets (\textit{asymmetric effect}), and has a long-lasting impact on its subsequent evolution (\textit{long memory}). SFs are intuitive and reflect human behaviors in response to risks. As shown later in our experiments, NN models that incorporate SFs have better accuracy compared to SOTA neural models. 

Our approach begins by studying the translation of established GARCH family econometric volatility models into their NN counterparts. The NN representations of the GARCH family that embed SFs then serve as building blocks plugged into the broader NN framework, aligning stochastic and NN approaches harmoniously. We call this approach the GARCH-NN approach. A major benefit is that the mathematical structures and statistical properties of the GARCH family models have been well studied with rigor by econometric researchers. The NN counterpart of the well-understood GARCH model can serve as the blueprint for building the final NN volatility model that can be understood and trusted by practitioners in the same way they trust the original stochastic models \cite{xai_2021}, helping to enhance NN volatility model's interpretability. 

Drawing from this perspective, we propose the GARCH-LSTM model as a motivative example for developing NN-based volatility models from stochastic volatility models. GARCH-LSTM seamlessly integrates GARCH's NN counterpart into the LSTM architecture. This design empowers the NN model not only to capture the stylized facts from GARCH but also to leverage LSTM's ability for long-short memory. Experiment results reveal that the amalgamation of the GARCH family (statistical) and LSTM (ML) models improves forecasting accuracy compared to their individual use. The contributions of our paper are listed as follows:

\begin{itemize}
    \item According to the authors' best knowledge, we are the first to study the equivalence relation between the classic GARCH family models in the econometrics field and NN in the ML field. This observation helps to bridge the gap between stochastic and ML approaches in financial volatility forecasting.
    \item Our approach constructs NN-based volatility models that integrate stochastic volatility models as building blocks. In this way, the stylized facts are directly infused into the NN framework, unlike traditional methods that simply take stochastic model outputs or parameters as input of the NN model. We develop the GARCH-LSTM model to demonstrate the superiority of our approach.
    \item We utilize five globally traded equities time series datasets and employ the GARCH-LSTM framework to compute the Value at Risk (VaR). Experiment results validate the existence of the GARCH-NN equivalence relation and combining the fundamental statistical (GARCH family models) and ML (LSTM) models yields improved results compared to employing each model in isolation.
\end{itemize}

\section{Related Work}
\label{sec:relatedwork}
\subsection{Stochastic Volatility Models}
The GARCH model \cite{BOLLERSLEV1986} is a widely used statistical model that captures volatility clustering and time-varying volatility in financial time series data. GARCH(1,1) is widely used in practice due to its simplicity and superior performance \cite{Bollerslev1987}. Many extensions have been developed after the pioneering work of GARCH. GJR-GARCH \cite{GJRgarch1993} aims to capture the asymmetry in the response of volatility, whereby negative shocks have a stronger impact compared to positive shocks. Early GARCH models were constrained by a short memory due to the exponential decay of the volatility weights.
Notably, FI-GARCH \cite{FIGARCH1996} and its extensions \cite{HGARCH2004,ST-FIGARCH2011} incorporate hyperbolic decaying coefficients, enabling these models to ``memorize'' long historical terms. MM-GARCH \cite{MuyiMixture2013} combines elements of both short-memory and long memory GARCH models. 

\subsection{Neural Network Volatility Forecasting Models}
Although the econometric time series models are advantageous in that they can be theoretically described based on statistical logic, they rely on various assumptions that do not align with the reality of highly dynamic and complex financial markets.
Over the past few years, significant advancements have been made in general-purpose deep learning (DL) models for time series forecasting, leveraging attentional mechanisms and transformer structures \cite{transformer2017,informer2021,kitaev2020reformer,DeepAR2020,N-BEATS2019,sen2019think,lai2018modeling,wu2021autoformer}.
Recent research has reworked linear models in time series forecasting and showcased their superiority in specific contexts \cite{zeng2022transformers, chen2023tsmixer}.
However, there exists evidence that DL struggles to outperform classical stochastic time series forecasting approaches \cite{Makridakis2018,elsayed2021really}, possibly due to that general-purpose NN models are not originally designed for the volatility forecast task to include unique stylized facts.

Attempts have been made to incorporate stylized facts into the ML model.
Most hybrid solutions take the outputs or parameters of stochastic models as features input to different types of NNs, like Multi-layer perceptron \cite{Khan2017-MLP,Pyo2018-mlp,Werner2016-mlp}, LSTM \cite{Kristjanpoller-ANN2014,LSTM-GARCH2020,rahimikia2020machine,LSTM-GARCH-Kim2018}, Transformer \cite{ramos2021multi}, and attention neural network \cite{SparseAttention2021,Zheng2019StockVP}. The ensemble approach combines the outputs from GARCH and NN \cite{Kakade-ensemble2022,Yan-hybrid2020}. \cite{DL-volatility-forecast-survey} systematically reviewed NN-based volatility forecasting. 
However, simply staggering two different models in a pipeline does not guarantee the interpretability or effectiveness of the approach.
Our paper distinguishes itself from existing NN-based volatility forecasting models in that we \textit{encode SFs into the NN structure}. The end-to-end GARCH-NN framework accelerates the model training and makes online forecasts feasible. Compared to sophisticated DL time series models, we intentionally choose basic NN models to build the fundamental equivalence relation between NNs and their stochastic GARCH counterparts. 
 

\section{Preliminaries and Problem Definition}\label{sec:problem_def}
The GARCH family has become the most popular way of parameterizing the dependence in volatility time series. The GARCH(1,1) model can be expressed in Equation \ref{equ:garch11}, where $r_t=log \frac{p_t}{p_{t-1}}$ denotes the log return at $t$, $\mathcal{D}$ stands for the distribution (typically assumed to be a normal or a leptokurtic one), $\{\epsilon_t\}$ may be observed directly, or it may be a residual sequence of an econometric model, $\psi_{t-1}$ denotes the historical information, $\gamma$ is a scalar, and $\theta(d)=\alpha \beta^{d-1}$ represents the contribution of $\epsilon_{t-d}^2$ in forecasting $\sigma_t^2$, which is an exponential time decayed function which dies out quickly with the increment of time lag $d$ due to the autoregressive term, leading to a short memory. $0<\alpha+\beta < 1$ guarantees the stationarity of the GARCH process.

\begin{equation}
\label{equ:garch11}
\begin{aligned}
r_t =& ~\mu_t + \epsilon_t, \quad \epsilon_t|\psi_{t-1} \sim \mathcal{D}(0, \sigma_t^2) \\
 \sigma^2_t =& \omega + \alpha * \epsilon^2_{t-1} + \beta * \sigma^2_{t-1} = \gamma + \sum\limits_{d=1}^{\infty} \theta_d \epsilon_{t-d}^2
\end{aligned}
\end{equation}

Given an input univariate time series with $l$ time steps, $\mathcal{E}_{t}^l = \{\epsilon_{t}, \epsilon_{t-1}, ..., \epsilon_{t-l+1} \}$, $l \leq L$, where $L$ denotes the entire time series length. According to the efficient market hypothesis, in practice, we treat $\mu_t=0$ and then $\epsilon_t = r_t$ in Equation \ref{equ:garch11}. The forecasting problem is defined below:
\begin{equation}
\label{equ:long_horizon_problem}
    \hat{\sigma}_{t+h}^2 = \mathcal{F} ( \mathcal{E}_{t}^l \vert \Theta_{\mathcal{F}} ) 
\end{equation}

\noindent where $\hat{\sigma}_{t+h}^2$ denotes the forecast future volatility at $t+h$, $h$ denotes the forecast horizon, $\mathcal{F}$ denotes the NN forecasting model, and $\Theta_{\mathcal{F}}$ refers to the model parameters. The goal is to design a highly interpretable model, meaning $\mathcal{F}$ either, has a well-studied NN structure whose property and performance are comprehensively studied and recognized or has equivalent mathematical counterparts having clear econometrics/statistical meaning, or both.

\section{Methodology}\label{sec:method}
The GARCH-NN approach of creating interpretable NN volatility models involves two key steps. Firstly, we establish an equivalence between GARCH models and their NN counterparts. Secondly, we seamlessly integrate the NN counterpart of GARCH into established NN blocks (such as LSTM), thereby ensuring the preservation of volatility stylized facts captured by GARCH within the NN framework.

\subsection{Equivalence Relation Between GARCH and NN}
\label{subsec:duality}
In this paper, the GARCH-NN equivalence is defined as both models sharing \textit{identical inputs, model structure, model parameters, loss function, and training processes}.

\subsubsection{Equivalence of Model Structure and Parameters}
\label{subsubsec:equ_structure}
Due to the page limit, we choose the three fundamental GARCH family models, namely, GARCH(1,1), GJR-GARCH, and FI-GARCH, corresponding to the volatility clustering, asymmetry, and long memory volatility stylized facts respectively. The NN counterparts of other GARCH family models can be established similarly.

\begin{figure}%
    \centering
    \subfloat[\centering Basic RNN Structure ]{{\includegraphics[width=0.45\linewidth]{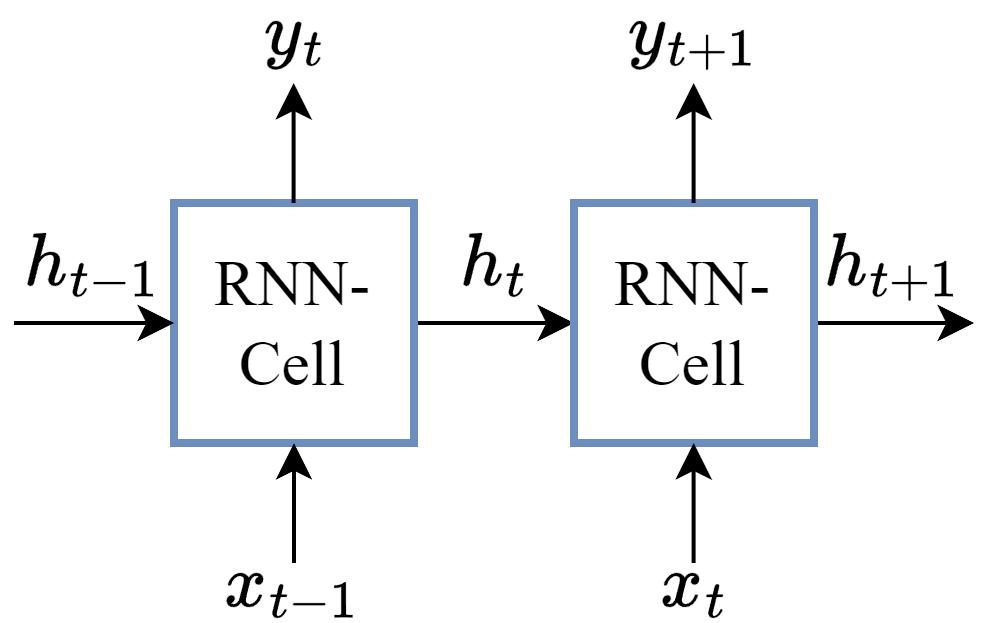}\label{subfig:rnn}}} %
    \subfloat[\centering GARCH(1,1)]{{\includegraphics[width=0.48\linewidth]{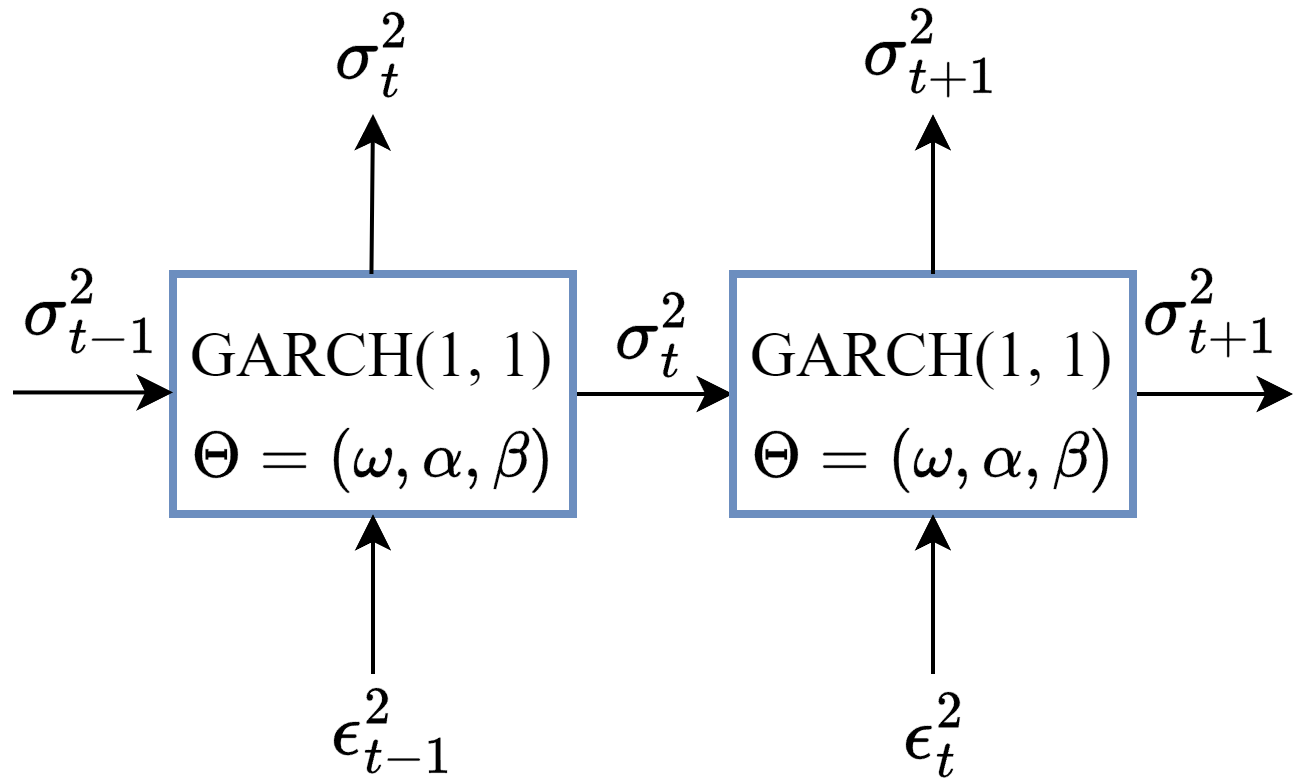}\label{subfig:garch11} }}
    \quad
    \subfloat[\centering GJR-GARCH]{{\includegraphics[width=0.48\linewidth]{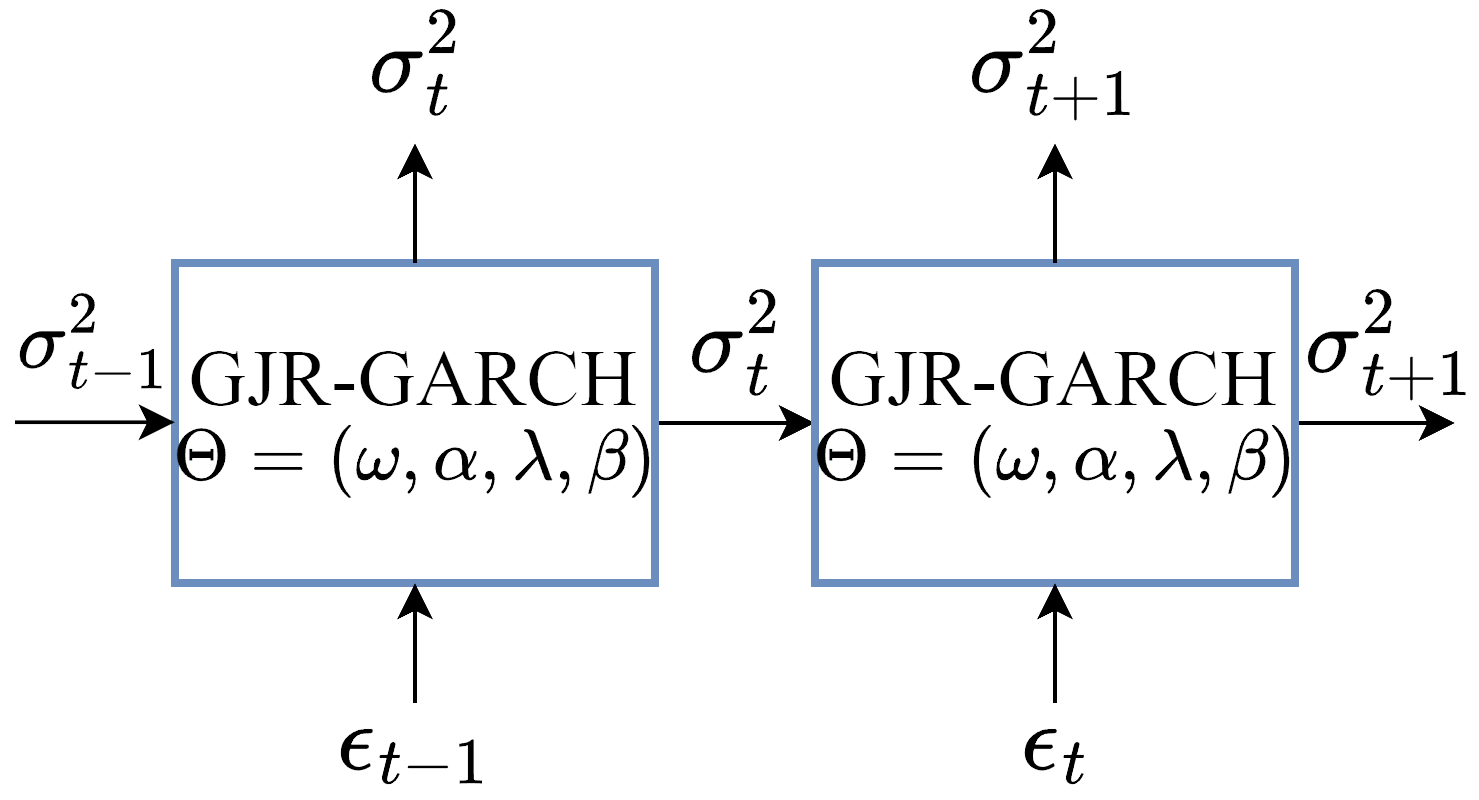}\label{subfig:gjr} }} %
    \subfloat[\centering FI-GARCH ]{{\includegraphics[width=0.45\linewidth]{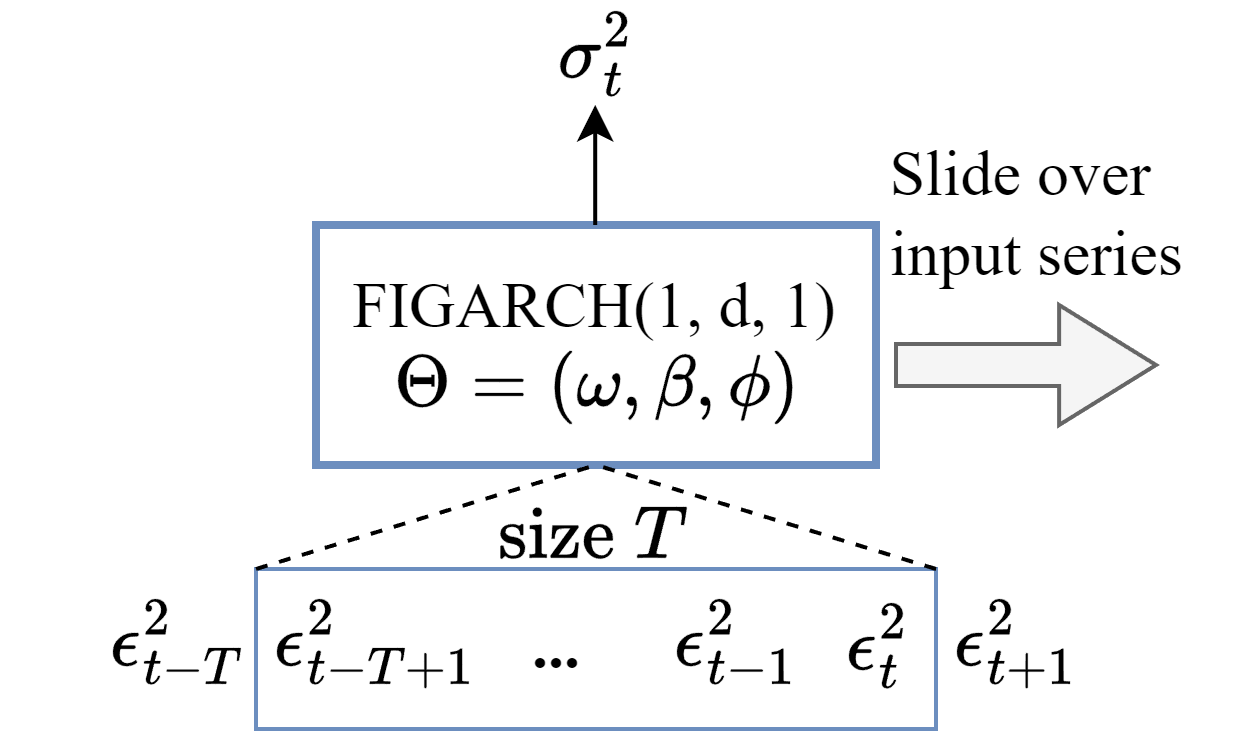}\label{subfig:figarch}}} %
    \caption{Equivalence between GARCH models and their NN counterparts.}%
\end{figure}

We start with GARCH(1,1), the simplest version of GARCH models. Figure \ref{subfig:garch11} illustrates the recursive structure of the GARCH(1,1) defined in Equation \ref{equ:garch11}, where the model input is $\sigma_{t-1}^2$ and $\epsilon_{t-1}^2$, and model output $\sigma_t^2$ is treated as the next GARCH cell's input. Intuitively, we can observe that it shares an identical recurrent structure as the RNN in Figure \ref{subfig:rnn}. Specifically, if we remove RNN's output layer and $tanh$ activation, and at the same time constrain both the input and hidden state to a scalar, the two models' structures are identical. Thus, theoretically, the GARCH(1,1) model is a special case of the RNN model with the truncation of the output layer and activation function. The RNN cell contains the parameter list $\Theta_{garch11}=(\omega,\alpha,\beta)$ and receives the observation list $X_{garch11}=(1,\epsilon_{t-1}^2,\sigma_{t-1}^2)$ and the output $\sigma_t^2$ is the linear combination $\Theta_{garch11} \cdot X_{garch11}$. Figure \ref{subfig:gjr} illustrates the RNN equivalence of the GJR-GARCH model. Compared to the GARCH(1,1) equivalence, the input to the RNN cell is $\epsilon_t$ instead of $\epsilon_t^2$ since the sign information of $\epsilon_t$ is required. The RNN cell contains the parameter list $\Theta_{gjr} = (\omega,\alpha,\lambda, \beta)$ and the observation list $X_{gjr}=(1,\epsilon_t^2,\mathcal{I}(\epsilon_t)*\epsilon_t^2, \sigma_{t-1}^2)$, where $\mathcal{I}$ is the sign function and the output $\sigma_t^2=\Theta_{gjr}\cdot X_{gjr}$.

In FI-GARCH(1, $d$, 1), the fractional $d$ controls the degree of long-memory behavior, and the parameter list $\Theta_{figarch} = (\omega, \beta, \phi)$ defines a decaying scheme for lags of $\epsilon_t^2$. The coefficient of $\epsilon_t^2 $ has the series expansion form as follows:
\begin{equation}
    \begin{aligned}
        \relax [1 - \frac{1-\phi B}{1-\beta B} (1-B)^d] \epsilon_t^2 = (\sum_{k=0}^{\infty} \lambda_kB^{k})\epsilon_t^2
        \approx \sum_{k=0}^{T-1}\lambda_k\epsilon_{t-k}^2
    \end{aligned}
\end{equation}
where $k$ denotes the lagging step, $B$ denotes the backshift operator (e.g. $B^k\epsilon_t^2 = \epsilon_{t-k}^2$), and each $\lambda_k$ is a function of $(\beta, \phi, d)$. In practice, a finite number of terms is obtained in the series expansion by using a truncation size $T$. Let $\Lambda = (\lambda_0, ..., \lambda_{T-1})$ be the weights computed based on $(\beta, \phi, d)$. Since $\lambda_k$ is independent with $t$, we can view $\Lambda$ as a sliding window over the $\{\epsilon_t^2\}$ series and this is similar to applying a 1-d convolutional operation to $\{\epsilon_t^2\}$. Thus, the long-memory nature of the FI-GARCH model can be represented by a basic Convolutional Neural Network (\textit{CNN}) structure, as illustrated in Figure \ref{subfig:figarch}. The kernel size is equivalent to the truncation size $T$ and the convolutional weights are $\Lambda$. The CNN cell slides over the $\{\epsilon_t^2\}$ series to perceive 
$X_{figarch} = \{\epsilon_{t-T+1}^2, ... \epsilon_t^2\}$ and $\sigma_t^2 = \Lambda \cdot X_{figarch}$. 

\subsubsection{Equivalent Loss Function}
\label{subsubsec:equ_loss}

While existing NN-based volatility forecast models commonly employ mean squared error (MSE) or mean absolute error (MAE) loss functions between predicted volatility $\hat{\sigma_t}$ and true volatility $\sigma_t$, this can present practical challenges to volatility forecast due to the inherent statistical nature of volatility and its various forms such as historical, implied, and realized volatility \cite{DL-volatility-forecast-survey}. To establish the GARCH-NN equivalence, we opt for the maximum likelihood approach as the loss function, a choice aligned with stochastic volatility models, as shown in Equation \ref{equ:obj}, where $N$ signifies the sequence length.

\begin{equation}
\underset{\Theta}{\arg \max } \Pi_{t=1}^N \mathcal{L} \left(\epsilon_t ; \Theta \right)=\underset{\Theta}{\arg \min }\sum_{t=1}^N -\log \mathcal{L}  \left(\epsilon_t ; \Theta\right) 
  \label{equ:obj}
\end{equation}

Assume $\epsilon_t$ in Equation \ref{equ:garch11} satisfies $\epsilon_t|\psi_{t-1} \sim \mathcal{N}(0, \hat{\sigma}_t^2; \Theta)$, then the negative log-likelihood is denoted in Equation \ref{equ:obj_normal}, termed as \textit{N-loss}.
\begin{equation}
\label{equ:obj_normal}
  -log \mathcal{L} \left(\epsilon_t;\Theta\right) = \frac{log \hat{\sigma}_t(\Theta)^2}{2} + \frac{\epsilon_t^2}{2\hat{\sigma}_t(\Theta)^2}
\end{equation}

We also adopt Student's $t$ density function in our experiments since studies \cite{LSTM-GARCH2020} have shown that it is suitable for volatility forecast. Then, $\epsilon_t$ satisfies $\epsilon_t|\psi_{t-1} \sim \mathcal{S}(0, \hat{\sigma}_t^2, v; \Theta)$ with a mean $\mu=0$, variance $\hat{\sigma}_t(\Theta)^2$ and degree of freedom $v$. The negative log-likelihood is then denoted as Equation \ref{equ:obj_stu}, termed as \textit{T-loss}.
\begin{equation}
\label{equ:obj_stu}
\begin{aligned}
-\log \mathcal{L} \left(\epsilon_t ; \Theta\right)= & \frac{\log \hat{\sigma}_t(\Theta)^2}{2} +\\
&\frac{v+1}{2} \log \left(1+\frac{\epsilon_t^2}{(v-2) \hat{\sigma}_t(\Theta)^2}\right) 
\end{aligned}
\end{equation}

The equivalence relation between the GARCH models and their NN counterparts can be established formally using identical model structures, parameters, and loss functions, while also feeding the same input and applying the same training settings, such as the SLSQP optimizer that is commonly used in sequential stochastic models. 

\subsection{GARCH-LSTM Model}
\label{subsec:lstm-duality}
LSTM is the classic sequential ML model whose capability to balance long and short historical information (memory) has been well studied and verified in various prediction tasks. Our goal is to design a model, named GARCH-LSTM, empowered with both the traditional GARCH family's capability of modeling the SFs and LSTM's ability to balance the long and short memory of volatility history. Equation \ref{equ:lstm_kernel} denotes the GARCH-LSTM model. Since there are many variations of the GARCH family models, we adopt a loosely coupled design that allows different GARCH models to easily plug into the LSTM framework. Specifically, in contrast to the conventional interpretation of LSTM's output gate $o_t$ as regulating the information selection from $c_t$, we reinterpret $o_t$ as the GARCH output and $c_t$ as the controller infusing LSTM's long and short-term memory effect. $\mathcal{K}_{garch}$ denotes the GARCH kernel function which can be flexibly replaced by different GARCH models' NN counterparts (e.g. GARCH(1,1), GJR-GARCH, FI-GARCH) introduced in the previous section. Notably, $\epsilon_{t-1}$ is the input to the model instead of $\epsilon_{t}$ which is originally used in LSTM since $\epsilon_t$ cannot be sampled before knowing $\sigma_t$, as shown in Equation \ref{equ:garch11}. We modify the output structure in the last line of Equation \ref{equ:lstm_kernel} by multiplying the output of the GARCH kernel function with $(1 + w*tanh(c_t))$. Influence from LSTM is ignored when $w = 0$ and the model is shrunk to GARCH's NN counterpart. With the increment of $w$, the LSTM module would have a greater impact on the final output by either magnifying $o_t$ ($c_t > 0$) or shrinking $o_t$ ($c_t<0$). GARCH-LSTM seamlessly integrates the GARCH family's NN counterparts encapsulating SFs into the LSTM framework. 

\begin{equation}
\label{equ:lstm_kernel}
\begin{aligned}
f_t & =\sigma_g(W_f * \epsilon_{t-1}+ U_f * \sigma_{t-1}^2 + b_f) \\
i_t & =\sigma_g(W_i * \epsilon_{t-1}+ U_i * \sigma_{t-1}^2 + b_i) \\
o_t & = \mathcal{K}_{garch}(\epsilon_{t-1}, \sigma_{t-1}^2;\Theta) \\
\tilde{c}_t & =\sigma_c(W_c * \epsilon_{t-1}+ U_c * \sigma_{t-1}^2 + b_c) \\
c_t & =f_t \odot c_{t-1}+i_t \odot \tilde{c}_t \\
\sigma_t^2 & =o_t \odot (1+ w * tanh\left(c_t\right))
\end{aligned}
\end{equation}

\section{Experiment}\label{sec:expriment}

\subsection{Experiment Settings}
\subsubsection{Datasets} 
We select five asset types covering stock indexes, exchange rates, and gold prices\footnote{Data were downloaded from https://finance.yahoo.com/}, which are widely traded equities by investors from all over the world. Following the common practice in the volatility forecasting literature \cite{bucci2020realized,bucci2017forecasting,andersen2003modeling}, we obtain the return series by computing the logarithm difference based on the daily close price series  $r_t=log \frac{p_t}{p_{t-1}}$, and generate the realized volatility based on the $k$-day average as $\sigma_t  = \sqrt{\sum_{i=0}^{k-1} \epsilon_{t-i}^2}$:
Here we set the window size $k=5$. The dataset is comprised of records in the form of $[(\epsilon_{t-k+1}, ..., \epsilon_t), \sigma_{t+h}^2]$. To avoid numeric underflow, we multiply the $\epsilon_t$ and $\sigma_{t+h}^2$ by a factor $100$. Table \ref{tab:datasets} summarizes the statistics of different datasets. ADF and p-value columns refer to the Augmented Dickey-Fuller test statistic and p-value. The near-zero p-value rejects the null hypothesis that a unit root is present in the time series data, indicating the stationarity of the time series. We also conducted the Kwiatkowski-Phillips-Schmidt-Shin (KPSS) test for each dataset, and the results failed to reject the null hypothesis that the time series is stationary.
We split the complete dataset into training, validation, and testing parts and the split ratio is roughly 8:1:1. We train and tune algorithms using the training and validation sets, respectively, and evaluate the model performance based on the testing set. 


\begin{table}[t]
\small
\centering
\begin{tabular}{c|ccccc}
\hline
Dataset & Length& Mean& Sd& ADF & P-value\\
\hline
S\&P 500& 2514&0.0414&1.078&-15.99&6.72e-29 \\
DJI& 2515&0.0314&1.93&-15.93&7.76e-29 \\
NASDAQ& 2516&0.0574&1.261&-13.66&1.56e-25 \\
EUR-USD& 2603&-0.0109&0.497&-22.08&0.0 \\
Gold& 2514&-0.0012&0.984&-51.95&0.0 \\
\hline
\end{tabular}
\caption{Summary of dataset statistics. }
\label{tab:datasets}
\end{table}

\subsubsection{Metrics}
Mean Squared Error (MSE) and Mean Absolute Error (MAE) between the ground truth volatility and the predicted volatility are used for performance evaluation. 

\begin{table}[t]
\small
\centering
\tabcolsep=0.12cm
\begin{tabular}{l|l||l|l}
\hline
\multicolumn{2}{c||}{\diagbox{Category}{Method}} & GARCH & NN Counterparts\\\hline\hline
\multirow{3}{*}{GARCH(1, 1)}&$\omega$&0.00359(0.00659)&0.00429(0.00659)\\ \cline{2-4}
&$\alpha$&0.00540(0.00596)&0.00543(0.00563)\\\cline{2-4}
&$\beta$&0.00923(0.0131)&0.0116(0.0161)\\\hline\hline
\multirow{4}{*}{GJR-GARCH}&$\omega$&0.00068(0.00073)&0.00081(0.00082)\\ \cline{2-4}
&$\alpha$&0.00073(0.00087)&0.00124(0.00119)\\\cline{2-4}
&$\beta$&0.00364(0.00291)&0.00415(0.00370)\\\cline{2-4}
&$\lambda$&0.00451(0.00748)&0.00363(0.00588)\\\hline\hline
\multirow{4}{*}{FI-GARCH}&$\omega$&0.164(0.367)& 0.0827(0.119)\\ \cline{2-4}
&$\beta$&0.0211(0.0201)&0.0363(0.0441)\\\cline{2-4}
&$\phi$&0.0189(0.0201)&0.0195(0.0375)\\\cline{2-4}
&$d$&0.0185(0.0282)&0.0249(0.0266)\\\hline
\end{tabular}
\caption{Parameter estimation comparison between GARCH models and their NN counterparts using simulation data.}
\label{tab:simulation_result}
\end{table}

\begin{table*}[t]
\small
\centering
\tabcolsep=0.12cm
\begin{tabular}{l|l||ll|ll|ll|ll|ll}
\hline
\multicolumn{2}{c||}{ \multirow{2}{*}{\diagbox{Method}{Dataset}}} &  \multicolumn{2}{c|}{S\&P 500} &\multicolumn{2}{c|}{DJI} &\multicolumn{2}{c|}{NASDAQ} &\multicolumn{2}{c|}{EUR-USD} &\multicolumn{2}{c}{Gold}\\
\cline{3-12}
\multicolumn{2}{c||}{}&MAE& MSE & MAE & MSE & MAE& MSE & MAE& MSE & MAE& MSE\\
\hline\hline
\multirow{2}{*}{GARCH(1,1)}&Vanilla&0.205&0.076&0.175&0.056&0.310&0.172&0.116&0.024&0.274&0.123\\\cline{2-12}
&NN Counterpart&0.221&0.086&0.191&0.065&0.313&0.173&0.116&0.027&0.267&0.117\\\hline
\multirow{2}{*}{GJR-GARCH}&Vanilla&0.207&0.074&0.190&0.060&0.296&0.157&0.114&0.023&0.278&0.126\\\cline{2-12}
&NN Counterpart&0.203&0.071&0.191&0.061&0.283&0.143&0.111&0.021&0.271&0.118\\\hline
\multirow{2}{*}{FI-GARCH}&Vanilla&0.213&0.076&0.183&0.057&0.305&0.170&0.128&0.029&0.279&0.125\\\cline{2-12}
&NN Counterpart&0.227&0.091&0.194&0.058&0.305&0.227&0.131&0.057&0.298&0.136\\\hline
\end{tabular}
\caption{Compare forecast performance with real-world dataset between stochastic models and their NN counterparts. }
\label{tab:ablation_study}
\end{table*}

\subsection{Validation of GARCH-NN Relation}
To validate the GARCH-NN equivalence relation in practice, we compare the training process, learned parameters, and model outputs (forecasting results) of GARCH models and their NN counterparts. 

\subsubsection{Training Process}
We implement the NN counterparts in PyTorch. The vanilla GARCH models use the SLSQP optimizer. Theoretically, the equivalence relation requires GARCH and its NN counterpart to have the same optimizer, while in practice, for better adaptability to NN architectures, we select the ADAM optimizer and use a dynamic learning rate scheduler which reduces the rate by a factor of $2$ if the validation performance stagnates. 
We use an early-stopping mechanism that terminates the training process if the model performance on the validation dataset does not improve for more than $20$ epochs. 
Both the vanilla GARCH models and their NN counterparts are trained by rolling over the training part of the time series and generating the out-of-sample predictions based on the same testing dataset. 

\subsubsection{Comparison of Learned Parameters with Simulation Data}
The close parameter estimation results of the stochastic GARCH models and their NN counterparts would indicate the validity of the equivalence relation. Since we do not have knowledge about the ground truth generation process of the real-world time series, we evaluate the parameter estimation results using simulation data. For each equivalence scenario, we generate 8 groups of simulated return $\{\epsilon_t\}$ series, each generated from the pre-fixed ground truth parameters ranging from (0.1,0.9) according to the corresponding GARCH process. The stochastic GARCH models and their NN counterparts are trained on the simulation data by the standard ARCH package \cite{arch} and PyTorch, respectively, and MSE is calculated between the estimated parameter values and the ground truth parameter values. 
Table \ref{tab:simulation_result} displays the average MSE and the standard deviation for each estimated parameter. 
From the experimental results, we can observe that both statistical and ML parameter estimation have low MSE compared to the original scale of parameters and this demonstrates the validity of the GARCH-NN equivalence relation. 
The experimental results also point out a way to automate the parameter estimation process of the GARCH family models, since as long as the NN counterpart is found, parameter estimation can be solved with backpropagation which saves great efforts for econometric models to develop explicit case-by-case parameter estimation algorithms.

\subsubsection{Comparison of Model Outputs}
We further compare the forecasting performance of the stochastic GARCH models and their NN counterparts using real-world time series data. If the forecasting performance is similar for both types of models, it would support the equivalence relation. Both vanilla GARCH models and their NN counterparts learn the model parameters based on the training and validation parts of the real-world datasets. Prediction is done in the testing dataset with horizon $h=1$. Table \ref{tab:ablation_study} shows the experiment results where ``vanilla'' denotes the stochastic volatility model using ARCH as the forecast package. Results show that the forecast performance of the stochastic models and their NN counterparts are close, which demonstrates the validity of the GARCN-NN equivalence relation. 

\begin{table*}[tb]
\centering
\small
\tabcolsep=0.12cm
\begin{tabular}{ll||ll|ll|ll|ll|ll|ll|ll|ll}
\hline
\multicolumn{2}{c||}{ \multirow{3}{*}{\diagbox[width=2cm]{Dataset}{Method}}} & \multicolumn{4}{c|}{MSE loss} &
\multicolumn{4}{c|}{N-Loss} & \multicolumn{4}{c|}{T-loss (v=3)} & \multicolumn{4}{c}{T-loss (v=5) }\\\cline{3-18}
&&\multicolumn{2}{c|}{RNN} & \multicolumn{2}{c|}{LSTM} &\multicolumn{2}{c|}{RNN} & \multicolumn{2}{c|}{LSTM} &\multicolumn{2}{c|}{RNN} & \multicolumn{2}{c|}{LSTM} &\multicolumn{2}{c|}{RNN} & \multicolumn{2}{c}{LSTM} \\\cline{3-18}
\multicolumn{2}{c||}{}&MAE&MSE&MAE&MSE&MAE&MSE&MAE&MSE&MAE&MSE&MAE&MSE&MAE&MSE&MAE&MSE\\\hline\hline
\multicolumn{2}{c||}{S\&P 500}&0.765&1.001&0.725&0.909&0.247&0.100&0.272&0.124&0.598&0.448&0.505&0.443&\textbf{0.200}&\textbf{0.068}&\textbf{0.216}&\textbf{0.087}\\\hline
\multicolumn{2}{c||}{DJI}&0.398&0.263&0.682&0.753&0.227&0.096&0.235&0.101&\textbf{0.184}&\textbf{0.050}&0.389&0.258&0.198&0.062&\textbf{0.182}&\textbf{0.059}\\\hline
\multicolumn{2}{c||}{NASDAQ}&0.748&0.952&1.017&1.723&0.461&0.369&0.442&0.385&\textbf{0.288}&\textbf{0.128}&0.805&1.142&0.432&0.336&\textbf{0.354}&\textbf{0.252}\\\hline
\multicolumn{2}{c||}{EUR-USD}&0.164&0.043&0.165&0.043&0.310&0.121&\textbf{0.132}&\textbf{0.027}&0.168&0.040&0.190&0.051&\textbf{0.156}&\textbf{0.038}&0.150&0.035\\\hline
\multicolumn{2}{c||}{Gold}&0.321&0.175&0.609&0.558&0.298&\textbf{0.143}&0.325&0.186&0.347&0.188&0.302&0.159&\textbf{0.298}&0.152&\textbf{0.293}&\textbf{0.147}\\\hline
\end{tabular}
\caption{Impact of different loss functions. The bold number indicates the best performance.}
\label{tab:qme_loss}
\end{table*}

\subsection{Model Evaluation and Analysis}\label{subsec:model_eval}
In this section, we evaluate and analyze the performance of the proposed framework together with the baseline models.  

\subsubsection{Impact of GARCH Loss Function}\label{subsec:loss_experiment}
Above we introduce the GARCH loss function N-loss and T-loss based on maximizing the likelihood of $\{\epsilon_t\}$ in Equation \ref{equ:garch11}. In this subsection, we compare their performances with MSE loss which is widely used in existing NN-based volatility forecast models. The experiment is conducted based on the classical RNN and LSTM models using the loss functions of MSE, N-loss, and T-loss (with different degrees of freedom). We compare their performances in forecasting the volatility values with forecast horizon $h=1$ (one day later) in the testing dataset. Table \ref{tab:qme_loss} shows the performance of different loss functions. We can see that N-loss and T-loss achieve significant performance enhancement compared with MSE loss in most datasets, demonstrating the superiority of the maximum likelihood-based loss function in the volatility forecasting task. The optimal performance is highlighted and we can see the T-loss with freedom (v=5) has the best performance in most cases, thus we select it as the loss function for training GARCH-LSTM in the following subsection.

\subsubsection{Overall Comparison}\label{subsec:overall}
To provide a comprehensive evaluation of the proposed framework, we compare the forecasting performance of the GARCH-LSTM model to both the state-of-the-art deep learning time series forecasting algorithms and the typical stochastic models with respect to different forecast horizons. 
Specifically, we include four transformer-based models, namely, \textit{Autoformer} \cite{wu2021autoformer}, \textit{Informer} \cite{informer2021}, \textit{Reformer} \cite{kitaev2020reformer}, and the original \textit{Transformer} \cite{transformer2017}, in the overall comparison. For these models, we input the returns series to the encoder and input the volatility series to the decoder to generate future volatility. 
We set the input length and label length to $126$ and use grid search to obtain the optimal hyperparameters. For GARCH-LSTM, we tune the initial learning rate to 1e-2 from the range [3e-2, 3e-4] and use GJR-GARCH as the kernel function.
To avoid potential information leaks, we ensure that all approaches are evaluated based on the same testing dataset and no testing data sample is used in the training stage. 

Table \ref{tab:overall_comparison} shows the overall comparison results concerning different horizon values.
Here, "1D", "3D", "1W", "2W", and "1M" refer, respectively, to the horizon value of one trading day ($h=1$), three days ($h=3$), one week ($h=5$), two weeks ($h=10$), and one month ($h=21$). We report the averaging results based on 10 runs with different random seeds.
We can observe that GARCH-LSTM achieves the best MAE and MSE in most dataset and horizon settings. Specifically, GARCH-LSTM has dominating performance in S\&P 500, NASDAQ, and Gold datasets, with around $3\%$ MAE and $10\%$ MSE improvement on average against the second-best approach. 
Most transformer-based methods do not perform well and this is aligned with the finding from \cite{zeng2022transformers}. The three stochastic models have steady performance although their comparison results may vary a bit due to the different characters of the five datasets. 
In general, most methods' forecasting capabilities decayed as horizons become larger and this is natural due to the uncertain and dynamic long-horizon future. In terms of stability and robustness, the proposed method generally exhibits smaller variances (\textasciitilde 1e-4) compared to deep learning models (\textasciitilde 1e-3).


\begin{table*}[tb]
\centering
\small
\tabcolsep=0.12cm
\begin{tabular}{l|l||ll|ll|ll|ll|ll|ll|ll|ll}
\hline
\multicolumn{2}{c||}{Methods} &\multicolumn{2}{c|}{GARCH-LSTM} &\multicolumn{2}{c|}{Autoformer} &\multicolumn{2}{c|}{Informer} &\multicolumn{2}{c|}{Reformer} &\multicolumn{2}{c|}{Transformer} &\multicolumn{2}{c|}{GARCH(1, 1)} &\multicolumn{2}{c|}{GJR-GARCH} &\multicolumn{2}{c}{FI-GARCH}\\
\hline
\multicolumn{2}{c||}{Metrics} & MAE & MSE & MAE& MSE & MAE& MSE & MAE& MSE & MAE& MSE & MAE & MSE & MAE & MSE & MAE& MSE\\
\hline\hline
\parbox[t]{2mm}{\multirow{5}{*}{\rotatebox[origin=c]{90}{S\&P 500}}}
&1D&\textbf{0.205}&\textbf{0.070}&0.636&0.546&0.494&0.387&0.489&0.381&0.494&0.386&\textbf{0.205}&0.076&0.207&0.074&0.213&0.076\\\cline{2-18}
&3D&\textbf{0.250}&\textbf{0.115}&0.576&0.483&0.497&0.391&0.497&0.395&0.497&0.390&0.276&0.136&0.258&0.121&0.253&0.117\\\cline{2-18}
&1W&\textbf{0.332}&\textbf{0.195}&0.606&0.542&0.497&0.391&0.497&0.394&0.497&0.390&0.358&0.225&0.346&0.212&0.336&0.198\\\cline{2-18}
&2W&\textbf{0.385}&\textbf{0.256}&0.571&0.488&0.491&0.386&0.490&0.387&0.491&0.385&0.413&0.288&0.410&0.293&0.399&0.273\\\cline{2-18}
&1M&\textbf{0.430}&\textbf{0.304}&0.567&0.523&0.485&0.381&0.485&0.386&0.485&0.380&0.441&0.326&0.439&0.316&0.430&\textbf{0.304}\\\hline\hline
\parbox[t]{2mm}{\multirow{5}{*}{\rotatebox[origin=c]{90}{DJI}}}
&1D&\textbf{0.170}&\textbf{0.056}&0.478&0.389&0.394&0.244&0.393&0.248&0.394&0.246&0.175&\textbf{0.056}&0.190&0.060&0.183&0.057\\\cline{2-18}
&3D&0.269&\textbf{0.107}&0.441&0.308&0.393&0.244&0.393&0.247&0.393&0.245&0.251&0.108&\textbf{0.250}&0.109&0.252&0.108\\\cline{2-18}
&1W&\textbf{0.336}&\textbf{0.185}&0.465&0.337&0.397&0.251&0.397&0.255&0.397&0.253&0.338&0.192&0.340&0.193&\textbf{0.336}&\textbf{0.185}\\\cline{2-18}
&2W&0.371&\textbf{0.219}&0.468&0.319&0.404&0.261&0.405&0.266&0.404&0.263&0.374&0.221&0.378&0.232&0.375&0.226\\\cline{2-18}
&1M&\textbf{0.412}&\textbf{0.272}&0.590&0.558&0.418&0.281&0.420&0.286&0.419&0.284&0.415&0.268&\textbf{0.412}&0.273&0.414&0.274\\\hline\hline
\parbox[t]{2mm}{\multirow{5}{*}{\rotatebox[origin=c]{90}{NASDAQ}}}
&1D&\textbf{0.288}&\textbf{0.144}&0.599&0.608&0.706&0.793&0.685&0.682&0.691&0.724&0.310&0.172&0.296&0.157&0.305&0.170\\\cline{2-18}
&3D&\textbf{0.320}&\textbf{0.198}&0.696&0.779&0.708&0.794&0.686&0.684&0.692&0.724&0.385&0.264&0.359&0.226&0.346&0.217\\\cline{2-18}
&1W&\textbf{0.421}&\textbf{0.309}&0.786&1.020&0.707&0.794&0.685&0.682&0.690&0.723&0.473&0.390&0.438&0.346&0.430&0.332\\\cline{2-18}
&2W&\textbf{0.498}&\textbf{0.420}&0.755&0.965&0.705&0.793&0.679&0.676&0.686&0.718&0.544&0.526&0.538&0.520&0.515&0.464\\\cline{2-18}
&1M&\textbf{0.567}&\textbf{0.571}&0.890&1.331&0.703&0.787&0.675&0.665&0.683&0.711&0.632&0.686&0.640&0.711&0.580&0.583\\\hline\hline
\parbox[t]{2mm}{\multirow{5}{*}{\rotatebox[origin=c]{90}{EUR-USD}}}
&1D&0.118&\textbf{0.023}&0.338&0.151&0.171&0.048&0.173&0.049&0.164&0.043&0.116&0.024&\textbf{0.114}&\textbf{0.023}&0.128&0.029\\\cline{2-18}
&3D&0.125&0.027&0.359&0.174&0.172&0.049&0.174&0.049&0.165&0.044&0.124&0.027&\textbf{0.123}&\textbf{0.026}&0.132&0.031\\\cline{2-18}
&1W&\textbf{0.131}&\textbf{0.030}&0.391&0.206&0.175&0.050&0.176&0.050&0.167&0.045&0.133&0.032&0.132&0.031&0.140&0.035\\\cline{2-18}
&2W&\textbf{0.136}&\textbf{0.033}&0.371&0.192&0.178&0.052&0.180&0.052&0.171&0.046&0.137&0.033&\textbf{0.136}&\textbf{0.033}&0.139&0.034\\\cline{2-18}
&1M&0.148&\textbf{0.039}&0.268&0.113&0.195&0.066&0.197&0.066&0.188&0.058&0.149&0.040&\textbf{0.147}&\textbf{0.039}&0.150&0.040\\\hline\hline
\parbox[t]{2mm}{\multirow{5}{*}{\rotatebox[origin=c]{90}{Gold}}}
&1D&\textbf{0.244}&\textbf{0.100}&0.684&0.595&0.360&0.200&0.350&0.191&0.369&0.207&0.274&0.123&0.278&0.126&0.279&0.125\\\cline{2-18}
&3D&\textbf{0.263}&\textbf{0.127}&0.744&0.713&0.361&0.200&0.347&0.189&0.369&0.207&0.299&0.149&0.302&0.152&0.300&0.150\\\cline{2-18}
&1W&\textbf{0.297}&\textbf{0.164}&0.774&0.778&0.362&0.200&0.349&0.190&0.370&0.207&0.323&0.175&0.326&0.179&0.330&0.181\\\cline{2-18}
&2W&\textbf{0.315}&\textbf{0.170}&0.758&0.735&0.362&0.201&0.349&0.190&0.370&0.208&0.318&0.172&0.321&0.175&0.320&0.177\\\cline{2-18}
&1M&\textbf{0.310}&\textbf{0.166}&0.462&0.341&0.355&0.194&0.344&0.185&0.362&0.200&0.311&\textbf{0.166}&0.311&0.168&0.317&0.174\\\hline
\end{tabular}
\caption{Overall comparison of approaches concerning multiple horizons. The bold number indicates the best performance.}
\label{tab:overall_comparison}
\end{table*}

\subsection{Application Study}
In the real-world financial industry, \textit{Value at Risk} (\textit{VaR}) is widely used in risk management. It seeks to measure market risks in terms of asset price volatility, which synthesizes the greatest (or worst) loss expected from a portfolio, within determined time periods and confidence intervals. Formally, VaR is defined for a long position in an asset $S$ over a time horizon $j$, with probability $p$ ($0 < p < 1$): 

\begin{equation}
    \label{equ:var}
    p = P(\Delta P_j \leq VaR) = F_j(VaR)
\end{equation}

\noindent where $\Delta P_j$ represents the gain or loss amount of position $P$, given by $\Delta P_j = |P_{t+j}-P_t|$ and $F_j(.)$ is the accumulated distribution function of the random variable $\Delta P_j$. For example, if a portfolio of stocks has a one-day 5\% VaR of \$1 million, that means that there is a 0.95 probability that the portfolio position changes $\Delta P_j$ will fall in value by less than \$1 million over a one-day period if there is no trading. To calculate the VaR it is necessary to have an estimate of the volatility of the asset's log returns for the analysis horizon. 

\begin{figure}[t]
\centering
\includegraphics[width=0.47\textwidth]{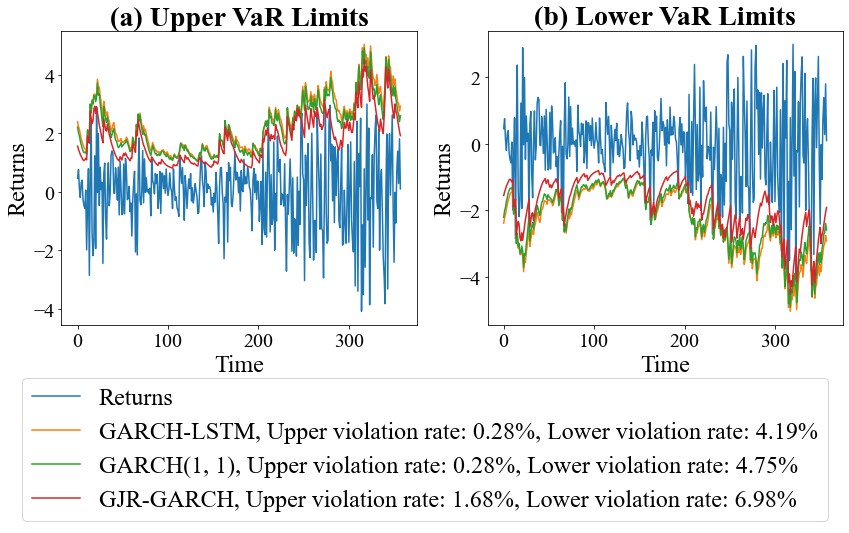}
\caption{VaR on NASDAQ dataset.}
\label{fig:case_study}
\end{figure}

The successful forecast of future volatility helps establish a reliable VaR, and in turn, the count of the number of times violating VaR would also help examine the volatility forecast performance \cite{var2007}. Thus, besides the MSE and MAE evaluation in Table \ref{tab:overall_comparison}, in this study, we count the violations of the VaR limits given by the number of excesses outside the confidence interval. The smaller violation rate indicates a better volatility forecast performance. The upper (lower) violation rate refers to the percentage that returns exceed the upper (lower) VaR limits over the total length of returns. In this study, we evaluate three different approaches, GARCH(1,1), GJR-GARCH, and GARCH-LSTM, that have top performances in Table \ref{tab:overall_comparison} based on the NASDAQ dataset.
Figures \ref{fig:case_study} (a) and (b) display the upper and lower violation rates of 5\% 1-day VaR of $1.65*\hat{\sigma}_t$ ($\mu_t=0$) in NASDAQ out-of-sample period. X-axis ranges from 2021-02-05 ($x=0$) to 2022-07-08 ($x=350$), and Y-axis denotes the return $\epsilon_t$. We can observe that all three approaches' total violation rates (upper+lower) are around 5\% which indicates the correctness of forecasting the volatility. We can also see GARCH(1,1) is a strong baseline which is consistent with the existing literature \cite{HansenPeter2005} and this explains why it is widely used in the real-world financial industry. It is also worth noting that this experiment favors approaches forecasting large $\hat{\sigma_t}$ since it leads to the large confidence interval that broadens the violation boundary and in turn loosens the violation criteria. Thus, the experiment results shown in Figure \ref{fig:case_study} should not be analyzed independently. Instead, it should be accompanied by the experiment results shown in Table \ref{tab:overall_comparison}.

\section{Conclusion}\label{sec:conc}

In this paper, we explore the equivalence relation between GARCH models and NN. Leveraging this relation, we propose a novel GARCH-NN approach for devising NN-based volatility models. This involves deriving the NN equivalents of GARCH family models, treating them as fundamental building blocks, and seamlessly integrating them into an established NN framework. This method allows the volatility stylized facts to be seamlessly infused into NN. We develop the GARCH-LSTM model to exemplify the GARCH-NN approach. Experiment results validate the GARCH-NN equivalence relation and show that combining the fundamental stochastic (GARCH family models) and NN (LSTM) models yields improved results compared to employing each model in isolation. For future work, we plan to extend the current research in the following directions: (1) We plan to explore if the GARCH-NN equivalence relation widely exists in the GARCH family beyond the three representative GARCH models mentioned in this paper to incorporate more SFs into the NN framework. 
(2) We plan to integrate the GARCH kernel into more NN frameworks and study if it would lead to better volatility modeling. 

\section*{Acknowledgements}
The research reported in this paper was supported in part by the Guangdong Provincial Key Laboratory of Interdisciplinary Research and Application for Data Science, BNU-HKBU United International College, project code 2022B1212010006, and in part by Guangdong Higher Education Upgrading Plan (2021-2025) of ``Rushing to the Top, Making Up Shortcomings and Strengthening Special Features'' with UIC research grant R0400001-22, Guangdong Higher Education Upgrading Plan (UIC-R0400024-21), Research Grants Council HKSAR GRF (No. 16215019). We appreciate Dr. Zhefang Zhou from BNU-HKBU United International College for insightful discussions and the anonymous reviewers for their helpful comments on the manuscript.

\bibliography{aaai24}

\end{document}